%% file: main.tex
\documentclass[preprint,journal]{vgtc}            

\onlineid{1634}

\vgtccategory{Research}

\title{A Design Space for Quantum Circuit Visualizations}

\author{%
  \authororcid{Hyeok Kim}{0000-0003-4340-4470} and 
  \authororcid{Leilani Battle}{0000-0003-3870-636X} 
}

\authorfooter{
  \item
  	Hyeok Kim is with the Korea Advanced Institute of Science \& Technology, and the work was done at the University of Washington.
    Leilani Battle is with the University of Washington.
  	E-mail: \{hyeokk, leibatt\}@uw.edu.
}

\input{sections/0-abstract}

\graphicspath{{figs/}{figures/}{pictures/}{images/}{./}} 

\usepackage{tabu}                      
\usepackage{booktabs}                  
\usepackage{lipsum}                    
\usepackage{mwe}                       
\usepackage{ccicons}                   

\usepackage{mathptmx}                  

\input{commands}

\begin{document}


\firstsection{Introduction}

\maketitle

\input{sections/1-intro}
\input{sections/2-related-work}
\input{sections/3-methods}
\input{sections/5-design-space}

\input{sections/5a-space-information}

\input{sections/5b-space-management}

\input{sections/5c-space-narrative}
\input{sections/6-1-examples}
\input{sections/6-discussion-1}
\input{sections/7-discussion-2}

\acknowledgments{%
  This work was supported in part by Google and the National Science Foundation (Award \# 2402718, \# 2514565, \# 2141506).%
}

\bibliographystyle{abbrv-doi-hyperref}

\bibliography{reference}








\end{document}

%% file: sections/0-abstract.tex
\abstract{%
    Quantum circuit visualizations play an essential role in supporting sense-making and communication of quantum programs. 
    While several tools exist for rendering quantum circuits, they vary widely in encoding options due to idiosyncrasies among machine and platform providers.
    We observe an opportunity to coalesce these disparate rendering approaches under a single, unified grammar to enable consistent, cross-platform enhancement of quantum circuit visualizations. 
    However, it is unclear how to design such a grammar to best support the quantum computing community. 
    Towards this end, we contribute a design space of quantum circuit visualizations by analyzing 182 static and 12 interactive cases collected from online tutorials and documentations, research publications, public presentations, and prior systems. 
    Based on our analysis, we discuss how our design space relates to existing visualization principles yet exhibits unique aspects. 
    We conclude with opportunities for future systems regarding data structure, cognition, and integrability.
}

\keywords{Quantum computing, quantum circuit visualization, quantum programming, visualization design space}

\teaser{
  \centering
  \includegraphics[alt={Seven quantum circuits labeled A through G are classified by spatiotemporal layout and format. There are three layout types: traditional, which uses a one-dimensional spatial layout and one-dimensional temporal layout; on-machine, which uses a two-dimensional spatial layout with no temporal dimension; and hybrid, which includes 3D, animated, and small multiples. There are two format types: circuit, which uses qubits and operations as symbols, and heatmap, which uses qubits and operations as cells. In subfigure A, a traditional circuit uses horizontal lines for qubits and overlaid symbols for operations. An overlaid box selects several operations with an annotation reading entangling. The bottom line is doubled, representing a classical register that stores measurement outcomes. In subfigure B, a traditional heatmap uses rows to represent qubits and columns to represent time progression, with cell color indicating the error rate. In subfigure C, an on-machine circuit displays qubits as dots on a grid, where grid lines indicate qubit connectivity. Several qubits are grouped with a highlighted box, and cells formed by grid lines are color-coded to represent stabilizer groupings. In subfigure D, an on-machine heatmap displays qubit connectivity across two sections for different gate types, with cell colors indicating gate fidelity. In subfigure E, a 3D circuit displays gates as 3D boxes along single qubit lines. In subfigure F, an animated circuit presents a sequence of animated on-machine circuit views. In subfigure G, a multiples heatmap consists of small multiples of on-machine heatmaps. Formats for subfigures E through G can be rendered as either heatmaps or circuits.},width=\linewidth]{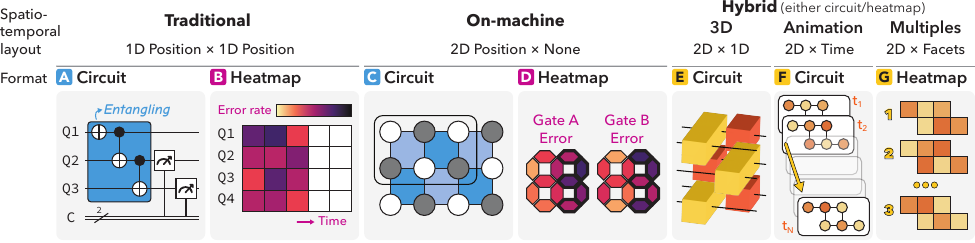}
  \caption{%
  	Example quantum circuit visualizations (\# indicates the identifier in our corpus). 
    (A \& B) Traditional visualizations with the vertical axis for qubits and the horizontal axis for the time order of operations:
    (A) Traditional circuit visualization where each line represents a qubit and shapes on those lines indicate operations (\#7 \& \#91). 
    (B) Traditional heatmap with each cell representing the quantified value of the corresponding qubit and time, such as error rate (\#136). 
    (C \& D) On-machine visualizations where qubits are laid out on a 2-dimensional position and no time information is given.
    (C) On-machine circuit diagram with qubit types and groupings (\#12 \& \#175). 
    (D) Two concatenated on-machine heatmaps with each cell representing the quantified variable and importance of each qubit pairing (\#90). 
    (E, F, \& G) Hybrid formats that add a time dimension on an 2-dimensional on-machine layout.
    (E) 3-dimensional traditional circuit (\#12).
    (F) An animated on-machine circuit (Patoka~\cite{kim2025:patoka}).
    (G) Small multiples of on-machine heatmaps where each sub-view stands for a time stamp (\#105).
    Images are redrawn from corresponding examples in our corpus for the presentation and copyright purposes.
    `\#\textit{N}' denotes the example index in our example gallery that is available via \url{https://see-mike-out.github.io/qc-circ-gallery/}.
    We further describe some of these examples in \Cref{sec:examples}.
  }
  \label{fig:teaser}
}

%% file: commands.tex
\usepackage{xspace}
\usepackage{listings}
\usepackage{array}
\usepackage{multirow}
\usepackage{soul}

\newcommand{\ie}{{i.e.,}\xspace}
\newcommand{\eg}{{e.g.,}\xspace}
\newcommand{\cf}{{c.f.}\xspace}
\newcommand{\ea}{{et~al.}\xspace}

\newcommand{\etc}{{etc.}\xspace}

\newcommand{\bpstart}[1]{\vspace{1mm} \noindent{\textbf{#1.}}}

\newcommand{\bstartnc}[1]{\vspace{1mm} \noindent{\textbf{#1}}}

\newcommand{\ket}[1]{$|{#1}\rangle$}

\newcommand{\ex}[1]{\##1}

\newcommand{\feature}[1]{\textbf{#1}}
\newcommand{\val}[1]{\textit{#1}}

%% file: sections/1-intro.tex
Visualizing quantum circuits plays an important role in communicating and teaching quantum information, such as reasoning about the structure of a quantum program, seeing how errors may propagate through a quantum machine, and illustrating abstract quantum computing concepts to lay-audiences~\cite{ashktorab2019:hqci,bethel2023:qcvis}.
Here, a \textit{circuit} mainly refers to a program (as a set of ordered operations). 
For example, quantum circuit diagrams (\eg~\Cref{fig:teaser}-A) are a widespread, standard method to represent quantum programs. 
Because processing quantum information depends on physical configuration, a circuit can be projected on the layout of a quantum processing unit (QPU or quantum chip), commonly expressed as on-machine layouts (\eg~\Cref{fig:teaser}-C \& D).

Currently, both academic and industry quantum communities have been successful in making medium-sized quantum programs and devices widely available. 
Yet, they have provided limited support for visualizing quantum programs and hardware-related information.
Those diagrams are mostly static, making it difficult to navigate larger programs; and other than conventional circuit diagrams tend to require ad-hoc scripting efforts~\cite{kim2025:patoka}.
We observe that these limitations stem from treating visualization as an offshoot of toolkits like Qiskit~\cite{qiskit} or Cirq~\cite{cirq}.
While recent visual analytic approaches~\cite{kim2025:patoka,lin2018:quflow,wen2024:quantivine,ruan2024:violet} propose interactive visualizations, such as filtering, and hierarchical aggregation, they also rely on specific technical stacks. 
Instead, treating visualization as a standalone toolkit that integrates with the current quantum computing ecosystem, we can better reason about encoding choices and incorporate recent advances in visualization and HCI. 

Therefore, we need a long-term vision for quantum circuit visualizations, ideally a grammar for supporting static and interactive versions that is agnostic to specific toolkits. 
A robust grammar requires a broad understanding of how current users are trying to visualize quantum circuits.
As a key first step toward this goal, our work aims at characterizing a design space for quantum circuit visualizations, with the following research questions:

\begin{itemize}
    \item RQ1. What constitutes quantum circuit visualizations?
    \item RQ2. What are the key design choices for quantum circuit visualization designs?
    \item RQ3. What are research opportunities for future work?
\end{itemize}

To answer these questions, we analyzed 182 static quantum circuit visualizations collected from computational notebooks of popular tutorials and documentations, notable research publications and conference presentations from the quantum computing and HCI communities, as well as 12 interactive visualizations introduced in prior work~\cite{kim2025:patoka,wen2024:quantivine,lin2018:quflow,ruan2024:violet,mcguffin2026:muqcs,ruan2024:quantumeyes,lamy2019:rainbow} and online tools~\cite{ibmComposer,quirk,crumble}.
We iteratively open-coded the static visualizations, and expanded our codes with the interactive visualizations.
We consolidate our analysis as a design space for quantum circuit visualizations (RQ1 \& RQ2).
Our design space is structured as \textbf{view-level} (\emph{position encodings and primary mark type choices}), \textbf{component-level} (\emph{core elements of quantum circuit visualizations}), \textbf{management-level} (\emph{how to select elements, encode additional properties, and manage data}), \textbf{narrative-level} (\emph{how to clarify key details and deliver high-level takeaways}), and \textbf{interaction-level} (\emph{supporting tasks around quantum circuits with interactivity}). 
We then reflect on our analysis by revisiting popular visualization design principles, including overview + detail~\cite{shneiderman2003:mantra}, multiple view consistency~\cite{qu2018:multiview}, and collaborative visualization~\cite{isenberg2011:collaborative}, to illuminate future research opportunities (RQ3). 
We conclude by discussing recommendations for technical work on quantum circuit visualization systems, in terms of data model, scalability, and integrability (RQ3).  In summary, this paper makes the following contributions:
\begin{itemize}
    \item We curate and qualitatively analyze 182 static and 12 interactive visualizations of quantum circuits, showcasing the breadth of encoding choices made by quantum computing users.
    \item Informed by our qualitative analysis, we introduce a design space for quantum circuit visualizations encompassing four aspects (view, information, narrative, and interaction).
    \item Based on our observations, we highlight open research questions for the visualization and quantum computing communities, including paths toward a platform-agnostic grammar for quantum circuit visualizations. 
\end{itemize}

%% file: sections/2-related-work.tex
\section{Background}\label{sec:rw}

The scope of our work is quantum circuit visualizations for digital quantum computing, which is a major paradigm in current quantum computing adopted by leading providers, such as IBM, Google, and Xanadu, as of 2026.
Below, we only describe key concepts relevant to our work. 
Readers can refer to Nielsen and Chuang~\cite{mike:and:ike} for theoretical backgrounds and Kim~\ea~\cite{kim2025:patoka} for a friendly introduction. 

\subsection{Quantum Circuits}

\begin{figure}
    \centering
    \includegraphics[alt={This figure is divided into two sections. Section 1 shows a Bloch sphere centered in a 3D Cartesian coordinate space defined by x, y, and z axes. A red vector extends from the origin to the surface, labeled as state Ket Psi. The azimuthal angle between the vector and the x-axis is labeled phi, and the polar angle between the vector and the z-axis is labeled theta. Section 2 shows four common quantum states illustrated using Bloch spheres. State Ket 0 points to the top pole along the z-axis, where theta is 0 and phi is 0. State Ket 1 points to the bottom pole along the z-axis, where theta is pi and phi is 0. State Ket plus points forward along the positive x-axis, where theta is half of pi and phi is 0. State Ket minus points backward along the negative x-axis, where theta is half of pi and phi is pi.},width=\linewidth]{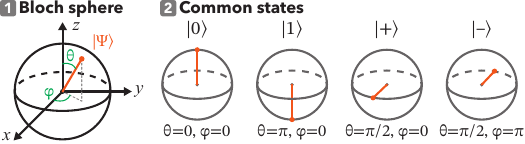}
    \caption{(A) The structure of a Bloch sphere. (B) Common states (\ket{0}, \ket{1}, \ket{+}, and \ket{-}) represented in the Bloch sphere format with their phase information ($\theta$, $\phi$).}
    \label{fig:bloch}
\end{figure}

For digital (gate-based) quantum computers, quantum circuits generally refer to quantum programs, essentially consisting of qubits, operations, and measurements. 
\textbf{Qubits} refer to information processing units, exhibiting superposed quantum states.
Simply put, the \textbf{state} of a qubit can be characterized as its probability ($P$) of being measured as 0 (or 1)\footnote{Measurement outcome depends on the physical device, but 0 and 1 are most common outcomes.}.
There are known states, such as \ket{0} ($P=1$), \ket{1} ($P=0$), \ket{+} ($P=0.5$), and \ket{-} ($P=0.5$). 
Here, \ket{X} (reads \textit{ket X}) is a notation for expressing quantum states.
The states of multiple qubits (or multi-qubit state) can be compressed into one expression, such as \ket{00101} or \ket{\Psi}.
A \textbf{Bloch sphere} is a common method to represent the state and phase of a qubit by showing it on the surface of a sphere (\Cref{fig:bloch}).
The same measured outcome can be from an infinite number of different underlying angular orientations (as depicted in the Bloch sphere), which is referred to as \textbf{phase}. 
Phase plays an essential role in interference that (de-)amplifies the likelihood of a certain measurement outcome.
For example, \ket{+} and \ket{-} have the same probability to be measured as 0, but they have different phases.
When another operation (\eg~Hadamard gate) is applied to them, they result in different states (\eg~\ket{0} and \ket{1}, respectively).

\textbf{Gates} (or operations) manipulate qubits to have a certain state. 
For example, a Hadamard gate (or H gate) turns \ket{0} to \ket{+}.
Common operations include H gate, NOT (or X) gate, and C-NOT/CX (controlled not) gate.
A C-NOT gate is a controlled gate that is applied to two qubits.
The target operation (NOT, in this case) is executed when the corresponding control qubit has the \ket{1} state.
For example, a C-NOT gate on a pair of qubits with \ket{10} (or \ket{11}) state (control-target) turns \ket{11} (or \ket{10}) whereas it would not affect \ket{00} or \ket{01}.
Some gates can have \textbf{parameters} for precise manipulation of qubits; for instance, an RX gate rotates a qubit by a specified angle.
A single gate or a set of multiple gates can always (and must) be expressed as a unitary matrix, a square matrix of which the conjugate transpose is its inverse\footnote{Essentially, given a unitary matrix $U$ and its conjugate transpose $U^*$, $U^*U$ = $UU^*$ = $I$, an identity matrix}.
Gates must be time-ordered for intended outcomes, and simultaneously executed gates are called a \textbf{moment} or \textbf{layer}.

Typically at the end of a circuit, qubits are \textbf{measured}. 
When measured, qubits lose their superposed states and become static values (\cf~basis states, eigenstates).
Thus, quantum computers run the same circuit many times to reconstruct the probability distribution of superposed states.
For \textit{quantum error correction} (detecting error and fixing it if needed), providers recently started supporting measurements in the middle of each run, which is often called \textit{mid-circuit measurement}.
When converting raw measurement outcomes into useful information or feeding mid-circuit measurement outcomes back to the quantum circuit (\eg~error correction), \textit{classical computing support} may be necessary. 
For example, quantum optimization algorithms employ classical optimization methods (\eg~gradient descent).

Programmers typically write a quantum circuit using logical qubits and operations.
Logical operations are those used in quantum information theory. 
Logical qubits tend to not have machine-related information, such as their on-chip position or fidelity.
Programmers need to convert \textbf{logical circuits} to physical circuits; this process is called \textbf{circuit optimization}, qubit mapping, or transpilation\footnote{These terms have slightly different meanings and nuances, the discussion of which is beyond our scope.}.
\textbf{Physical circuits} have qubits with their machine-related information and operations that are decomposed into some basic physical operations.
Different machines support different basic operations.
Physical qubit locations are important because the more distant two physical qubits are, the more error-prone gates between them are. 
Plus, many error correction codes (\eg~surface code, lattice surgery) depend on qubit locations.

\begin{figure}
    \centering
    \includegraphics[alt={This figure compares classical and quantum program and architecture visualizations. For program visualizations, classical program visualizations collapse bits after each evaluation, whereas quantum visualizations do not. Classical visualizations offer higher layout flexibility, whereas quantum visualizations are more constrained due to a fixed temporal dimension. For architecture diagrams, classical architecture diagrams use abstract granularities with flexible connectivity across diverse unit elements. In contrast, quantum architecture diagrams are typically defined at the strict qubit level with fixed connectivities and uniform properties per qubit.},width=\linewidth]{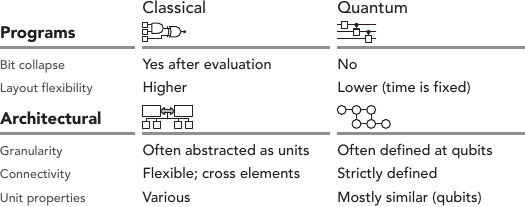}
    \caption{Comparing quantum circuit visualizations to classical circuit visualizations. This table only summarizes common design tendencies.}
    \label{fig:cmp-classical}
\end{figure}

\subsection{Visualizations for Quantum Circuits}

Quantum circuits are commonly expressed as visualizations for programming and communication purposes. 
Many popular toolkits like Qiskit~\cite{qiskit}, Cirq~\cite{cirq}, and PennyLane~\cite{pennylane} offer methods for creating static quantum circuits. 
Traditional formats as in \Cref{fig:teaser}-A express qubits as horizontal lines and operations as symbols and boxes on top of the lines corresponding to the applied qubits. 
A major constraint in quantum circuit visualizations is that there are three essential dimensions, horizontal and vertical locations of qubits and the time order of operations per qubit.
Plus, larger quantum circuit visualizations may produce illegible views.
During analysis, we also observed that advanced circuit visualizations like heatmaps often need ad-hoc scripts, extra toolkits, and custom visual edit.

Quantum circuit visualizations are different from classical ones fundamentally, often stemming from how they are conceptualized.
As compared in \Cref{fig:cmp-classical}, qubits are never merged into a single bit unlike classical bits, so qubit lines tend to stay until the end of a circuit
with limited flexibility in repositioning edges (as in flow charts).
Next, while classical architectures include various types of abstract units that can be flexibly connected (\eg~RAM, control logic), qubits tend to have strict connectivity constraints.
These unique aspects of quantum circuits motivate tailored visualization techniques.

Recently, more HCI and visualization-based approaches introduced more dynamic quantum circuit representations.
For example, beginners can use quirk~\cite{quirk} and IBM Quantum Composer~\cite{ibmComposer} to interactively compose quantum circuits via drag-and-drop interactions, yet it is challenging to manually drag hundreds of gates.

Crumble~\cite{gidney2021:stim,crumble} offers interactive on-machine diagram and traditional views for composing error correction codes. 
Patoka~\cite{kim2025:patoka} provides animated on-machine diagrams to show progression of a quantum program; and an interactive cross-highlighting visualization to allow for comparing logical and physical circuits for debugging purposes.
To mitigate scalability concerns, Quantivine~\cite{wen2024:quantivine} supports an overview given a large circuit and filtering operations and qubits.
QuFlow~\cite{lin2018:quflow} shows how parameters of operations have downstream influences to operations afterwards.

In quantum computing practices, toolkits are typically specific to machine providers as they mediate machines to programming environments. 
For instance, Qiskit is primarily designed for IBM machines, and some providers have adapted versions of Qiskit (\eg~IonQ\footnote{\url{https://docs.ionq.com/sdks/qiskit}}) for their machines. 
For the same reason, the above visualization tools are strongly tied to specific technical stacks, preventing those useful techniques across different settings. 
As open-sourced software ecosystem is important in making quantum computing more widely adoptable~\cite{Bova2025}, platform-agnostic quantum circuit visualization systems are necessary.
To provide a concrete foundation for such tools, we contribute a design space for quantum circuit visualizations.
Based on our analysis, we further provide recommendations for future tools.

\begin{figure}
    \centering
    \includegraphics[alt={This figure compares four spatiotemporal visualization types: maps, traditional quantum circuits, on-machine quantum circuits, and network visualizations. For element placement, maps and on-machine circuits feature fixed spatial locations. Networks arrange elements algorithmically by default. Traditional quantum circuits offer flexible qubit positions, but operation positions are strictly determined. For dimensions, maps, on-machine circuits, and networks use two dimensions for space and zero for time, whereas traditional circuits use one dimension for space and one for time. For element sets, maps, quantum circuits, and networks typically use fixed element sets, though networks can support dynamic sets and quantum circuits can vary focal elements over time. For hierarchy and connections, element hierarchies are inherently defined in maps, but must be explicitly specified in quantum circuits and networks. Edges represent predefined geographic connections in maps, qubit lines in traditional circuits, hardware connectivity in on-machine circuits, and measured relationships in networks.},width=\linewidth]{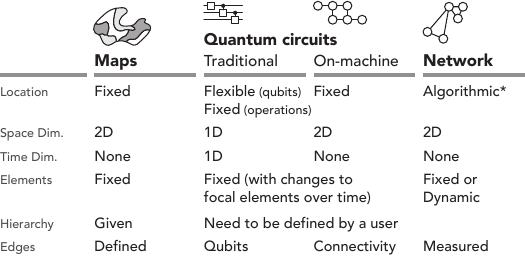}
    \caption{Comparing quantum circuit diagrams to spatiotemporal map and network visualizations. This table only means tendencies, assuming a single, static, simple view. *When no geospatial location is given.}
    \label{fig:spatiotemporal}
\end{figure}

\subsection{Visualization with Spatiotemporal Constraints}

Quantum circuits can be characterized as spatiotemporal data, considering qubit locations as 2D space and moments as time.
There are other visualization domains sharing the spatiotemporal and scalability constraints: time-serial maps and networks.
We briefly compare quantum circuit visualizations to those methods, as summarized in \Cref{fig:spatiotemporal}.
First, element positions tend to be given or fixed for maps and on-machine circuits; algorithmically decided for networks; and adjustable for traditional circuits (qubit lines).
For traditional circuits, given that the time dimension already occupies a position encoding, it is challenging to represent physical or functional proximity of qubits by projecting a 2D qubit grid on a 1D axis.
Next, over a time dimension, essential map elements and qubits do not disappear nor are they added, whereas these changes may still occur for networks. 
For quantum circuits, while important qubits (\eg~under operation) may differ at a time, they still exist and occupy physical locations that cannot be easily ignored.
In addition, maps tend to have (semi-) natural hierarchy (\eg~country-province-city) as a baseline\footnote{This does not mean that there is a fixed, universal one.}, quantum circuits need explicit, custom definitions.
Lastly, edges tend to be defined for maps (\eg~bus routes); measured for networks; given for qubits on traditional circuits; and fixed for on-machine diagrams (qubit connectivity).

To better motivate future research directions later in discussion, we briefly describe popular techniques used for representing spatiotemporal methods and their perceptual effectiveness, according to prior work~\cite{kjellin2008:2d3d,mota2023:urban,yang2019:spatio,adrienko2003:spatio,gelernter2016:spatio}.
High-level techniques include small multiples, animation, and 3D space-time cubes.
Small multiples for spatiotemporal visualizations consist of static views for each timestamp, while animations show them over time.
3D space-time cubes add another position axis to represent time.
Perception studies tend to report that 2D small multiples are more perceptually effective than 3D~\cite{tittle1995:distortion,norman1996:length,todd2004:shape,kjellin2008:2d3d,mota2023:urban} and animated views~\cite{kjellin2008:2d3d,mota2023:urban}.

%% file: sections/3-methods.tex
\begin{figure}
    \centering
    \includegraphics[alt={This figure features four sections detailing the distribution of a corpus comprising 182 static quantum circuit visualizations. Section A focuses on sources. Visualizations originate primarily from research papers and documentation for Qiskit, PennyLane, and Cirq, with additional examples from conference materials. Section B covers creation methods. Half of the visualizations were generated programmatically using Qiskit, Cirq, PennyLane, Python, or LaTeX, while the other half were custom-built. Section C details scale. The corpus predominantly features circuits with 5 to 30 gates and 5 to 20 qubits, alongside larger cases containing over 300 gates and 100 to 200 qubits. Section D shows topics. The topic distribution covers error-related circuits at 28 percent, theoretical circuits at 15 percent, and transpilation-related circuits at 8 percent, application circuits at 40 percent, including search, optimization, natural sciences, cryptography, and quantum machine learning. Approximately 9 percent have no specific topic.},width=\linewidth]{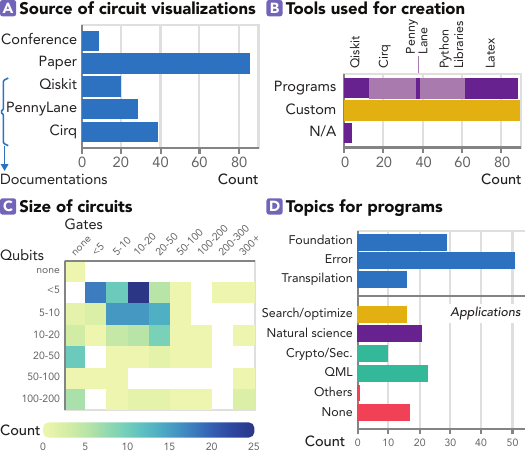}
    \caption{The distribution of our 182 static visualization sample in terms of source, creation tools, circuit size, and problem topic.}
    \label{fig:dist}
\end{figure}

\begin{figure*}
    \centering
    \includegraphics[alt={Our design space consists of five high-level design decisions from view, component, management, narrative, to interaction. First, the view level addresses abstraction, composition, layout, and format. Second, the component level details element information, visual representation, and design considerations. Third, the management level governs selection, visual encoding, references like axes and legends, data transformations, and space management techniques. Fourth, the narrative level encompasses titles, annotations, visual emphasis, and structural attachments. Fifth, the interaction level encompasses exploration and composition interactions.},width=\linewidth]{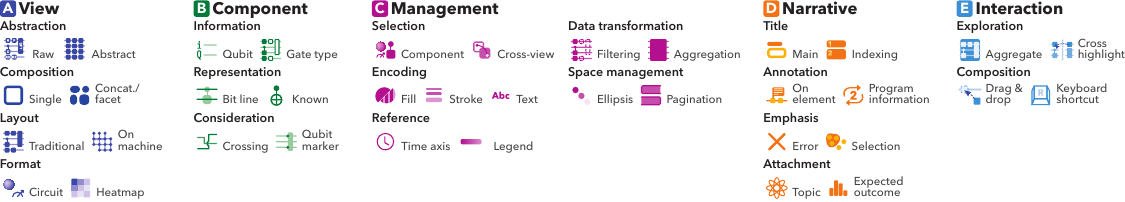}
    \caption{We characterize our design space in terms of view (A), component (B), management (C), narrative (D), and interaction (E).
    }
    \label{fig:overview}
\end{figure*}

\section{Methods}\label{sec:methods}

We analyzed 182 static and 12 interactive circuit visualizations from academic and practical sources using open-coding methods.

\subsection{Data Collection}

\bpstart{Static visualizations}
While we do not claim a comprehensive corpus of quantum circuit visualizations, we aimed to collect cases that can show a diverse range of characteristics.
In total, we collected 182 static circuit visualizations from tutorial notebooks of major quantum computer providers (IBM, Xanadu-PennyLane, and Google), conferences for researchers and developers (\eg~IEEE Quantum Computing \& Engineering, IBM's Quantum Developer Conference), and Nature's flagship journals (published after January 2025).
We stopped collecting examples as we reached a point where we were not adding any new open-coding tags (theoretical saturation).
The sample list and collection details are provided in the Supplementary Material.

\bpstart{Interactive visualizations}
We complemented our design space with 12 interactive circuit visualization cases. 
Currently, interactive quantum circuit visualizations are supported by specific tools, so we surveyed prior approaches, including: (1) circuit optimization checker and (2) result viewer from Patoka~\cite{kim2025:patoka}, (3) parameter propagation chart from QuFlow~\cite{lin2018:quflow}, (4) quantum machine learning circuit from VIOLET~\cite{ruan2024:violet}, (5) aggregated circuit visualization of Quantivine~\cite{wen2024:quantivine}, (6) Quirk composer~\cite{quirk}, (7) Circuit Composer~\cite{ibmComposer} and (8) Machine Explorer from IBM Quantum Platform, (9) Crumble for error correction~\cite{crumble}, (10) QuantumEyes~\cite{ruan2024:quantumeyes}, (11) Muqcs for state simulation~\cite{mcguffin2026:muqcs}, and (12) rainbow boxes by Lamy~\cite{lamy2019:rainbow}.

\subsection{Analysis}

We started with open-coding of the collected visualizations with no prior principles. 
Our initial codes had a form of key-value pairs. 
For example, a key for `Format' took several values, such as `traditional', `on-machine-full', `on-machine-partial', `on-machine-heatmap', and `heatmap'. 
We modified codes as we collected more cases and iterated over the design space.
For example, we separated the code for `Format$\rightarrow$on-machine-heatmap' from `Format$\rightarrow$heatmap' because they contained different spatial information (2D physical qubits vs. 1D qubit lists).
This distinction later informed the characterization of `Layout' and `Format' design choices.

We iterated over the codes in three rounds of review to minimize coding mistakes and inconsistencies. 
First, we collected all the keys in our codes and used them to inspect if each case in our sample was missing any information for those keys. 
Second, when there were systematic missing information, we filled them with some default values. 
For example, cases for traditional circuit diagrams were often missing information about qubit lines. Then, we added `Qubit line$\rightarrow$standard' codes for those cases.
We did this in a conservative way so that we did not affect any previously assigned codes. 
Third, we took the co-occurrence information of the key-value pairs. 
For example, `Format$\rightarrow$traditional' co-occurred `Qubit line$\rightarrow$standard' highly frequently (more than 80\% of the time). 
We looked at cases that did not have such co-occurrence to see if there were any mistakes.

Once we completed our cleaning steps, we had over 400 open codes.
We consolidated these codes into a design space in an iterative way, which involved another pass over our corpus.
We used prior surveys in relevant areas such as narrative~\cite{segel2010:narrative}, genomic~\cite{nusrat2019:genomic}, responsive~\cite{kim2021:responsive}, geospatial~\cite{schottler2021:geonetwork} and dashboard~\cite{bach2023:dashboard,srinivasan2025:zoo,javed2012:composite} visualizations to inform the structure of our design space for quantum circuit visualizations.

\begin{figure*}[t]
    \centering
    \includegraphics[alt={This figure consists of five sections, numbered from 1 to 5. The first section describes an initial design with a traditional circuit format. This circuit has 5 qubits, 15 gates, and 3 measurements. Below the circuit diagram, a bar chart shows the aggregated fidelity information for each moment. The second section shows view- and component-level changes to the initial design, such as (a) an on-machine diagram, (b) fidelity information mapped to node color, (c) special glyph designs for different gate types, which result in multiple scenes. Then, the scene progression can be done via (d) small multiples or (e) animation. Alternatively, (f) a traditional heatmap format can be used to depict disaggregated fidelity information. The third section shows management-level changes to the original circuit, including (a) filtering and (b) aggregation. The fourth section shows narrative-level changes, featuring annotations and emphasis for program-related information. Lastly, the fifth section displays interaction-level changes, such as selection features and tooltips providing detailed information like state vectors and probabilities.},width=\linewidth]{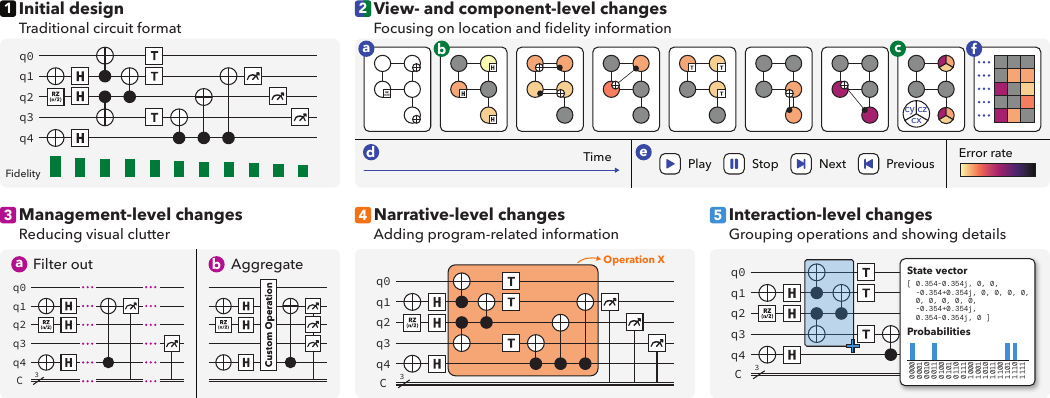}
    \caption{Given an initial design with a traditional layout and a circuit format (1), a user can make alternative view-level choices (on-machine layout) so as to display on-machine locations of qubits (2a). In addition, the user can add qubit fidelity information, such as error rate (\ie~additional components) while preserving gate representations (2b) or replacing them with special glyphs (2c). To show the time dimension, the user can use a grid composition with a time axis (2d) or a (self-paced) animation (hybrid layout) (2e). Alternatively, the user can use a traditional heatmap format (2f). When the circuit becomes too cluttered,
    the user can filter out (3a) or aggregate (3b) information. To emphasize a certain group of components (\eg~operations), the user can highlight them and a related annotation (4). Lastly, to show more detailed information, such as a state vector or expected probabilities, the user can insert a brush interaction and tooltip (5).}
    \label{fig:running}
\end{figure*}

%% file: sections/5-design-space.tex
\section{A Design Space for Quantum Circuit Visualizations}

Through qualitative analysis of our collected examples, \textbf{we contribute a design space for quantum circuit visualizations}, as shown in \Cref{fig:overview}.
The primary goals of our design space are to inform design choices in making quantum circuit visualizations and to provide vocabularies for describing them (RQ 1\&2).
We complement our proposed design space with an interactive design gallery in the Supplementary Material.

\subsection{Overview: Design Choices}

We describe our design space in terms of design choices that a creator would consider in creating quantum circuit visualizations, as illustrated in \Cref{fig:running}.
First, a creator would need to decide on \textbf{view-level} properties, such as the level of abstraction (guided by purpose), view composition, spatiotemporal layout, and overall format (as compared in \Cref{fig:running}-1 and \Cref{fig:running}-2).
Next, the creator needs to represent the core \textbf{components} (qubits, gates, and measures) of a circuit and their characteristics (as compared in \Cref{fig:running}-2a and 2b).
To allow for incorporating additional information (\eg~fidelity values) and adjust visual complexity and enhance presentability, the creator would need to \textbf{manage} circuit visualizations by making selections, encoding properties, transforming data, and organizing space (\Cref{fig:running}-3).
Lastly, the creator may want to add \textbf{narrative} elements (\Cref{fig:running}-4) for communication and \textbf{interactions} for exploration and creation (\Cref{fig:running}-5).

We characterize our design space in terms of design choices because higher level decisions often shape subsequent design choices at large.
For example, when using a heatmap format, then qubits are represented as cell rows for circuit layouts (\Cref{fig:running}-2f) and cells for on-machine layouts (\Cref{fig:teaser}D; view $\rightarrow$ component).
Yet, they are not always deterministic; for example, creators can overlay traditional gate representations on top of on-machine layout (\Cref{fig:running}-2, \ex{11}, \ex{168}) or attach fidelity metrics to a traditional circuit (\Cref{fig:running}-1, Patoka~\cite{kim2025:patoka}).
Similarly, components types and representations influence how they are selected and subsequently their narrative roles within the final visualization (component $\rightarrow$ management $\rightarrow$ narrative). 

\bpstart{Data Assumptions}
While it is incorporated in the component-level, we briefly summarize common data types and properties in quantum circuits.
Qubits and operations (including gates) are considered essential quantum circuit elements along with their properties.
Qubit properties include, but are not limited to, fidelity and connectivity; and operations properties include fidelity, applied qubits, parameters, \etc
Qubits and operations also have properties that evolve through a program, such as superposed states (in complex matrices), amplitudes, density, entanglement, \etc

\begin{figure}[t]
    \centering
    \includegraphics[alt={There are three subsections, numbered from 1 to 3, displaying different composition and progression methods. The first section shows a grid composition where heatmaps are laid out on a grid defined by two parameters. The second section shows a circuit-in-circuit composition, where a child circuit is embedded within the main circuit. The third section shows an overview and detail composition with a zoom-lens progression, where three circuits are concatenated using left-to-right arrows, and the detailed techniques behind the second circuit are rendered below and connected using a zoom-lens format.},width=\linewidth]{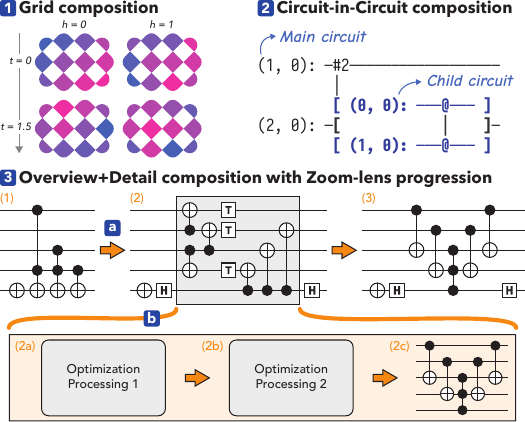}
    \caption{Composition and progression (view-level) examples. (1) A grid composition (redrawn from \ex{106}). (2) A circuit-in-circuit composition in a text mode (redrawn from \ex{75}).
    (3) A case with multiple composition and progression methods (redrawn from Case \ex{113}): (a) horizontal concatenation with arrow-based sequencing; and (b) overview+detail composition connected by a zoom-lens look.}
    \label{fig:case113}
\end{figure}

\subsection{View-level}\label{sec:view}
As a specialized domain, quantum circuit visualizations have commonly used high-level structural, or \textit{view-level}, choices, such as the level of \textit{abstraction}, \textit{layout}, and \textit{format}.

\bpstart{Abstraction}
Circuit visualizations have different levels of abstraction, in terms of details about building algorithms or procedures represented in them.
We frequently observed cases where the communication purpose of a visualization guides its level of abstraction.
First, \feature{programming visualizations} have fine-grained circuit information.
Programmers often produce them to inspect their circuit codes (\ie~communication with themselves or fellow programmers). 
Next, \feature{presentation} has medium to low abstraction, edited for communication purposes, such as public talks or research papers. 
With profound background knowledge (\eg~obtained by reading a paper), a programmer can have a good sense of how to build represented contents.
Next, \feature{illustrations} have \textit{highly abstract} information mostly as an indication of circuit components within a bigger picture like error mitigation algorithms (\ex{131}). 
Illustrations do not include replication information, but they give a conceptual snapshot of an algorithm or procedure. 

\bpstart{Composition and Progression}
Unless showing a \val{single-view} visualization, a creator should choose the \feature{composition} methods for multiple views.
Similar to \Cref{fig:teaser}D, views can be \val{faceted} or \val{concatenated}. 
Next, when there are certain criteria (\eg~different parameter values, time progression), sub-views can be laid out on a \val{grid} (\Cref{fig:running}-2, \Cref{fig:case113}-1, \ex{105}, \ex{73}).
When an operation consists of multiple gates, programmers can have the high-level operation and add another view for the details, laid on top of the circuit (\val{circuit-in-circuit}; \Cref{fig:case113}-2, \ex{85}) or shown separately (\val{overview+detail}; \Cref{fig:case113}-3, \ex{57}).

Creators can show how views are related (\feature{progression}).
Concatenated views often use \val{symbols}, such as arrows for orders (\ex{113}, \Cref{fig:case113}-3a, \Cref{fig:narrative}-1) and equality signs for transformation methods (\ex{29}). 
Overview+detail compositions can have \val{zoom-lens} looks (\Cref{fig:case113}-3b, \ex{70}).
These composition and progression methods can be combined.
For example, \ex{113} has concatenation and overview+detail compositions and symbol and zoom lens progressions (\Cref{fig:case113}-3).

\bstartnc{Layout} refers to how space (qubit locations) and time (moments) coordinates are positioned or timed.
\val{Traditional} layouts (\eg~\Cref{fig:teaser}-A \& B, \Cref{fig:running}-1) have one space dimension for the order of qubits and one time dimension for the moments. 
When there is an underlying physical qubit coordinates (\eg~physical programs), 2-dimensional locations need to be projected to a 1-dimensional list.
These traditional layouts can have \val{horizontal} or \val{vertical} \feature{directions}.
Next, \val{on-machine} layouts (\eg~\Cref{fig:teaser}-C \& D, \Cref{fig:running}-2) represent qubits as their relative 2-dimensional locations on a quantum chip.
Circuit visualizations tend to use this layout to deliver physical information.
As a stand-alone, on-machine layouts inherently lose a time dimension.
Albeit rare, \val{node-edge} layouts have qubits and gates as nodes and show their relationships (\eg~applied qubits, next gates) as edges.
Node-edge layouts may help with visual clutters, yet the applicability is limited.
For example, it might become difficult to tell qubits affected by gates (\ex{167}).

Because these layout options sacrifice either space or time dimension, there are \val{hybrid} approaches to compensate for the lost information. 
First, \val{3D} (\eg~\Cref{fig:teaser}-E) layouts complement traditional layouts by using 2-dimensional locations instead of 1-dimensional projection while preserving the time axis.
Next, \val{animations} (\eg~\Cref{fig:teaser}-F) show the progression of an on-machine circuit over time within the same view. 
The timing of an animation can be static (changing scenes by a fixed certain time interval, Crumble~\cite{crumble}), dynamic (based on the time required for processing a moment, Patoka~\cite{kim2025:patoka}), or self-paced (similar to interactive slideshows, Patoka~\cite{kim2025:patoka}).
We classify animations as a layout option (instead of a progression method) because this method is used typically to show changes to the properties of the same elements (qubits/gates) in our corpus, unlike other visualization animations that can show different data variables.
Lastly, \val{small multiples} (\eg~\Cref{fig:teaser}-G) lay out the temporal changes to an on-machine circuit view, often using a grid composition.

\bstartnc{Format} refers to representation methods for core components like qubits and gates.
\val{Circuit} formats show qubits and gates as themselves, using bit lines, dots, and gate notations (\eg~\Cref{fig:teaser}-A, C, E, \& F).
To support terminal-like environments, toolkits often support \val{text}-based representations (\feature{mode}) of circuit formats in addition to \val{graphical} ones.
On the other hand, \val{heatmap} formats represent qubits or gates as colored cells, frequently to include fidelity-related information (\eg~error rate, decoherence time). 
For example, \Cref{fig:teaser}-B represents qubits as rows but their gates as cells to show the gate error rates with a color encoding, and \Cref{fig:teaser}-D shows qubit connectivity (between two qubits) as cells to show their error rates.
Plus, \Cref{fig:teaser}-G lays out qubits as cells to show similar fidelity information.

%% file: sections/5a-space-information.tex
\begin{figure}
    \centering
    \includegraphics[alt={There are three sections, numbered 1 to 3, showing different methods for dynamic bit lines in traditional circuit formats. The first section shows qubit markers that indicate the qubits affected by an operation spanning multiple bit lines, as well as the crossing of bit lines. The second section shows bit lines starting and ending at different moments. The third section shows bit lines merging and diverging along the circuit.},width=\linewidth]{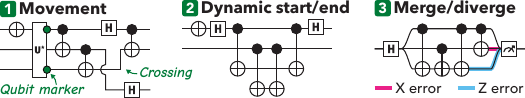}
    \caption{Dynamicity in bit lines (component-level): (1) Movement of bit lines (redrawn from \ex{114}, \ex{165}). Multi-qubit gates ($U^*$) can have qubit markers to indicate to which qubit they apply to. (2) Dynamic start and end (redrawn from \ex{156}). (3) Merged and diverged lines (redrawn from \ex{171}). Additional information (\eg~error propagation of Crumble~\cite{crumble}) can be \textit{projected} on the shades of qubit lines.}
    \label{fig:bitline}
\end{figure}

\subsection{Component-level}\label{sec:info-1}
Quantum circuit visualizations have core components to display, representing computation units, such as qubits, gates, and measurement.
They tend to occupy the position channels, which is usually decided by format and layout choices.
We characterize these components in terms of conveyed \textbf{information}, visual \textbf{representation} methods, and related \textbf{considerations}. 
For readability, we describe them by each computational unit.
Because there are a wide variety of possible components, we only brush over popular and unconventional choices. 
Readers can refer to our Supplementary Material for the full details. 

\bpstart{Qubit}
Because quantum programs and related procedures are defined at qubits, it is important to show qubits in circuit visualizations. 
Key \feature{information} regarding qubits includes name, index, the \val{number} of qubits, \val{type}/\val{group}, \val{state}, \val{fidelity}, and \val{mapping from logical to physical} qubits. 
Fidelity information includes error rates and decoherence times (after which the qubit loses its state information).

Layout and formats often decide qubit \feature{representations}. 
For example, traditional layouts tend to show qubits as vertical positions using \val{single lines} for circuit formats (\Cref{fig:teaser}-A, \Cref{fig:running}-1) and \val{rows} for heatmap formats (\Cref{fig:teaser}-B, \Cref{fig:running}-2f).
On the other hand, on-machine circuits show qubits as \val{dots} (\Cref{fig:teaser}-C, \Cref{fig:running}-2) and on-machine heatmaps represent them as \val{cells} (\Cref{fig:teaser}-G).
In displaying qubits, there are several \feature{considerations}. 
Creators can \val{group} qubits in terms of their role in a program or procedure. 
Common distinctions include data qubits storing information vs. ancilla qubits assisting operations (\ex{124}) and different stabilizers for error correction purposes (\ex{175}, \Cref{fig:teaser}-C).
As opposed to static bit lines in \Cref{fig:running}-1, bit lines can have \val{dynamicity}, as illustrated in \Cref{fig:bitline}.
For example, creators can move relevant bit lines closer to each other, optionally indicating \val{line crossing} (\Cref{fig:bitline}-1, \ex{114}).
Some circuits try to reduce visual clutter by showing bit lines \val{partially}  (\Cref{fig:bitline}-2, \ex{156}), \val{omitting} them (\ex{11}), or \val{merging} them (\Cref{fig:bitline}-3, \ex{174}, \ex{60}). 

\bpstart{Qubit connectivity}
Because how qubits are connected on a chip is necessary for procedures like error correction and circuit optimization, on-machine visualizations often encode qubit connectivity. 
Key \feature{information} includes \val{2-qubit} connectivity (\eg~2-qubit gates) and \val{N-qubit} connectivity (stabilizers for error correction, typically four qubits).
2-qubit connectivity is often \feature{represented} as \val{grid edges} for circuit formats (\Cref{fig:teaser}-C), using edge color, width, and dashes to encode groups, error rates, \etc, and \val{cells} in heatmap formats (\Cref{fig:teaser}-D). 
To show different 2-qubit gate errors, some circuits use \val{custom glyphs} (\Cref{fig:running}-2c, \ex{69}, \ex{70}).
N-qubit connectivity is represented as \val{cells} shaped by 2-qubit grid edges (\eg~blue rectangles in \Cref{fig:teaser}-C).

\begin{figure}
    \centering
    \includegraphics[alt={There are two sections showing variations for gates and classical procedures. The first section shows C-NOT gate notations drawn on top of an on-machine heatmap. The second section shows two methods, labeled 2-a and 2-b, for classical procedures. Subsection 2-a shows measurement notations on two qubits directly connected to an error-correction operation applied to the remaining qubits. Subsection 2-b shows measurement notations connected to a procedure and then to a classical optimizer, which feeds back to the beginning of the original circuit.},width=\linewidth]{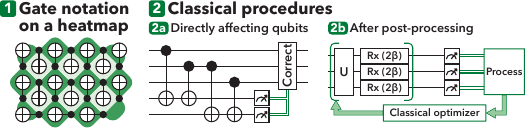}
    \caption{Gate and classical procedure representation examples (component-level). 
    (1) Using known gate notations (for C-NOT) on a heatmap (redrawn from \ex{187}). (2) Classical procedures: (2a) directly affecting qubits (\eg~error correction, redrawn from \ex{48}) and (2b) applying classical post-processing (redrawn from \ex{99}) 
    }
    \label{fig:gates}
\end{figure}

\bpstart{Gate}
The \val{types} of gates (\eg~C-NOT, H) are core \feature{information} for quantum programs.
On-machine heatmaps tend to display fidelity information, so gate types are often \textbf{given} (\ex{89}).
To \feature{represent} them, traditional circuit formats use widely \val{known notations} (\eg~$\bigoplus$ for NOT gates, and $\bullet-\bigoplus$ for C-NOT gates as in \Cref{fig:teaser}-A).
Although rare, it is also possible to use those notations on top of a heatmap format (\Cref{fig:gates}-1, \ex{187}).
Some visualizations use \val{custom symbols} (\Cref{fig:running}-2c, \ex{70}).
For precise control or optimization problems, it may be important to show gate \textit{parameters} (\Cref{fig:gates}-2b, QuFlow~\cite{lin2018:quflow}). 

Comprehensibility and gate behaviors motivate some design \feature{considerations}.
For example, visualizations for presentation or communication tend to \val{group} low-level gates and merge them as a big box operator with names or signs.
Such merged (or multi-qubit in general) operators may appear across multiple qubits.
To clarify to which qubits they apply, a creator can add some \val{qubit markers} (\Cref{fig:bitline}-1). 
We observed one case (\ex{34}) that represented the \val{pulse} data of gates (how a laser device works to execute gates) using a wave line.

\bpstart{Measurement and classical computing support}
Qubits can be measured at the end or in the middle of processing, and quantum algorithms often utilize the measured outcomes by processing them via classical computing support, such as optimization or error detection decoder.
Qubits are measured along with a \val{Pauli axis}, with the Z axis (or computational axis) being most common (measuring 0 or 1). 
Otherwise (X or Y), circuits need to indicate that (\ex{182}, \ex{160}).
Some circuits show the \val{expected values} of measured outcomes (\ex{63})
or the \val{probability} of measuring 0 or 1 (\Cref{fig:running}-5, \ex{107}).
Creators can choose to add a \val{classical bit line} (often double line) to indicate a classical register (\Cref{fig:teaser}-A). 
Quantum programs can feed the measurement outcomes back to other qubits.
In this case, the feedback methods, such as \val{directly affecting qubits} (\Cref{fig:gates}-2a, \ex{48}) or via classical \val{post-processing} (\Cref{fig:gates}-2b, \ex{99}) are key information.
Some creators add \val{attachments} (images or diagrams) via post-processing methods to illustrate final or mid-circuit measurement outcomes (\ex{154}, \ex{99}).

%% file: sections/5b-space-management.tex
\subsection{Management-level}\label{sec:info-2}

Once deciding what core components to show and how to show them, creators often transform them and encode additional information.
For example, components can be aggregated to indicate their roles or meanings in a circuit (\Cref{fig:running}-3b), or they can have additional information like fidelity and measurement probabilities shown via additional encodings (as a subsequent change in \Cref{fig:running}-2b). 
Doing so requires several abilities.
At baseline, creators should be able to \textbf{select} different components.
Creators should also be able to \textbf{encode} properties to such selections along with \textbf{references} like axes and legends.
To highlight semantics and reduce visual complexities, creators may want to \textbf{transform} circuit data (\eg~aggregating, filtering) with visual techniques for \textbf{space management}.

\bpstart{Selection}
Circuit visualizations make selections at different \textbf{levels of units}, such as \val{components} (as described in \Cref{sec:info-1}), \val{moments} (time orders of operations), parts of \val{circuits}, distinctions between quantum and classical processing (\val{QPU/CPU}), and a subset of \val{views}. 
\feature{To indicate} such selections, a creator can choose to use text \val{annotations}, visual \val{emphases}, \val{visually merging} elements, and using visual \val{encodings}.
For example, common aggregation strategies include merging multiple gates into a larger gate or collapsing multiple classical lines into one.
Elements can be selected \val{across different views} to show correspondence in difference before and after applying circuit modification methods (\ex{34}) or relevance to topic problem (\eg~Case \ex{168} maps gate groups to atomic energy levels).
Patoka's circuit optimization viewer~\cite{kim2025:patoka} supports interactive cross-view selection via hover highlight.
\val{Nested} selections are also common (\ex{17}, \ex{154}).

\bpstart{Encoding}
Components can have additional properties, such as qubit fidelity, gate types, and gate groups. 
Because space and time dimensions already occupy position encodings, those additional properties are represented as visual channels of corresponding \val{objects}.
For example, the on-machine circuit in \Cref{fig:teaser}-C has qubits positioned based on their on-chip locations while qubit fill colors indicate their roles.
When an object needs to encode more information (\eg~qubit type + qubit group), it is also common to add a \val{shade} behind it (\Cref{fig:bitline}-3) or \val{text} on top of it. 
For instance, as illustrated in \Cref{fig:bitline}-3, a qubit line can have shade to indicate how a gate error on a qubit propagates throughout the circuit, with its color denoting the type of error.
We observed common visual \feature{channels} \val{fill} color/pattern, \val{stroke} color/width/dash, \val{shape}, and \val{text}.
For animated cases (\eg~\Cref{fig:running}-2d, \cite{kim2025:patoka}), the operation time of a gate can be directly mapped to the \textit{timing} of each scene.

\bpstart{References}
For reference purposes, circuit visualizations use \feature{axes} for \val{space} (\eg~qubit location) and \val{time}. 
For non-positional encodings, we observed cases using \val{quantitative} (\Cref{fig:teaser}-B) and \val{categorical} (\Cref{fig:bitline}-3) \feature{legends}.
To provide detailed information, circuit visualizations can use legends for child gates for an aggregated gate (\ex{147)} and qubit layout grid as a legend (\ex{15}).

\begin{figure}
    \centering
    \includegraphics[alt={This figure consists of three sections, numbered 1 to 3, depicting space management techniques. The first section shows a portion of a traditional circuit visualization filtered out or erased, which is indicated using long tilde lines. The second section shows operations and bit lines filtered out from a traditional circuit visualization, where certain multi-qubit operations point to their corresponding qubits that have been omitted. The third section shows the pagination of an extremely long original circuit split across multiple pages.},width=\linewidth]{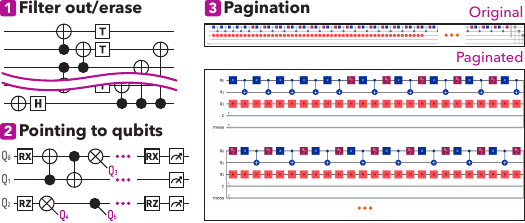}
    \caption{Space management examples. (1) Filtering out or erasing a part of a circuit (redrawn from \ex{184}). (2) Removing qubit lines and pointing to related qubits (redrawn from \ex{65}, \ex{5}, Quantivine~\cite{wen2024:quantivine}). (3) Pagination of a large-scale circuit (redrawn from \ex{7}, \ex{37}).}
    \label{fig:space-manage}
\end{figure}

\bpstart{Data transformation and space management}
Quantum circuits often include repeated components (\eg~applying the same set of gates multiple times) or less important elements (\eg~ancilla qubits), increasing visual complexity.
Quantum circuits can employ data transformation and space management techinques to deal with visual complexity.
First, filtering and aggregation were the most common data transformation techniques in our corpus.
\val{Filtering out} repeated or unimportant components (\Cref{fig:running}-3a) can be done systematically (\eg~every other qubits) or through customization (\eg~specific qubits).
Quantum circuits can show how operations form higher-level operations (\eg~Quantum Fourier Transform) or sub-routines by \val{aggregating} low-level operations into a single box in a static (\Cref{fig:running}-3b, \Cref{fig:bitline}-1) or interactive (Quantivine~\cite{wen2024:quantivine}) way.
Next, visual space management techniques can be categorized as omission indication and pagination.
Omitted components can be indicated using  \val{ellipsis} (\Cref{fig:space-manage}-1) or \val{indexing} (of omitted qubits, \Cref{fig:space-manage}-2).
Quantum software toolkits allow for \val{slicing the circuit into multiple pages} (\Cref{fig:space-manage}-3) to reduce the extensive graphical density of large-scale circuits (Qiskit) in a static or interactive (Patoka~\cite{kim2025:patoka}) way.

%% file: sections/5c-space-narrative.tex
\begin{figure}
    \centering
    \includegraphics[alt={This figure has two sections showing narrative elements. The first section shows how narrative elements indicate changes to a circuit. The original on-machine circuit features several qubits that are grouped, linked, and labeled with numbers. An arrow symbol with an annotation then connects the original circuit to a new circuit. This case demonstrates the use of titles for subviews, annotations for indexing specific qubits and marking view progression, and emphasis for qubit selection. The second section shows how annotations and emphasis indicate program information. On a traditional circuit diagram, spark symbols labeled with the letter E are placed on bit lines to indicate error locations, and a portion of the circuit is highlighted using a dashed box annotated as repeated M times. This case demonstrates the use of annotations for program information, such as loops, and emphasis for qubit selection and error locations.},width=\linewidth]{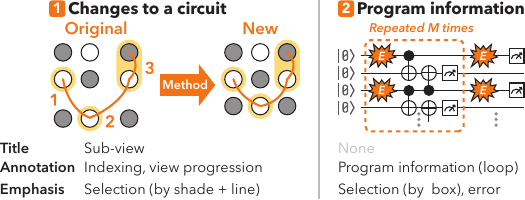}
    \caption{Narrative-level examples. (1) Concatenated views with their relationship and sub-view titles (redrawn from \ex{168}, \ex{170}). (2) Program information (loop, error location) on a circuit (redrawn from \ex{123}).
    }
    \label{fig:narrative}
\end{figure}

\subsection{Narrative-level}

As we surveyed circuit visualizations from research papers and conference presentations, we observed a wide variety of narrative elements, such as annotation and emphasis.
Creators made them by using custom graphic tools from scratch,
creating a baseline diagram via toolkits and decorating them with custom tools like Powerpoint\footnote{We could tell them based on icons and color palettes of Powerpoint.}, or using Latex with custom functions. 
Yet, narrative elements are not solely for presentation purposes; they are also essential for delivering program-related information (\eg~error simulation, loops). 

\bpstart{Title}
Circuit visualizations have \val{titles} to indicate their topics or major insights. 
Faceted and concatenated views have \val{sub-view title} to inform how those views are different (\eg~error rate for different gates), as shown in \Cref{fig:teaser}-D and \Cref{fig:narrative}-1.
They can have \val{indices} as well for ordering and referencing purposes, as they appear in figure captions.

\bpstart{Annotation and emphasis}
To identify circuit components and indicate their functional roles, quantum circuits use text annotations and visual emphasis.
They include \val{indices}, \val{names}, and \val{group markers} to \val{program information} and view \val{progression} details.
For example, \Cref{fig:narrative}-1 has indices, shade, connection lines for selected qubits and view progression (``method'' on an arrow).
\Cref{fig:narrative}-2 includes a dash-stroked box that groups a section of a circuit, which is annotated for the number of repetition of that section. 
It also includes the time and location of errors happening while running the circuit (``E'' in explosion symbols).
There are other program information, such as the mapping between logical and physical qubits after circuit optimization.

\bpstart{Attachment}
We observed that quantum circuits are attached with additional information for contextualization, classical processing methods, summary statistics, and expected results. 
First, to present related \val{topics}, such as a node-edge graph for a search problem (\ex{20}), atoms for physics problems (\ex{111}), and a neural network pictogram (\ex{158}).
Next, circuit visualizations visually display algorithms or procedures for \val{classical processing} (\eg~circuit optimization), as shown in \Cref{fig:case113}. 
Third, circuits can have \val{summary statistics} (\ex{10}, \ex{76}) such as the number of different gates, the number of qubits, \etc
While it does not have a visualization, case \ex{44} shows a detailed list of summary statistics. 
Lastly, quantum circuits can include \val{expected results}, such as measurement outcome distribution (\ex{124}), a list of simulated outcomes (\ex{101}), and superposed qubit states (\ex{76}).
VIOLET~\cite{ruan2024:violet} has an attached sub-view about the expected performance of a QML algorithm.

\begin{figure}
    \centering
    \includegraphics[alt={This figure has two sections showing different interaction methods. The first section shows interactive cross-highlighting. There is a logical circuit, labeled 1-a, and a physical circuit, labeled 1-b. The logical circuit includes a Hadamard gate and a C-NOT gate, with the CNOT gate highlighted using shading. In the physical circuit, the corresponding portion for the C-NOT gate is similarly highlighted, and the two highlighted parts are connected by an arrow line as if there is an hover interaction. The second section shows interactive circuit composition, divided into two subsections labeled 2-a and 2-b. Subsection 2-a shows dragging and dropping an operation from a gate panel onto a traditional circuit diagram, which updates qubit state visualizations and an outcome distribution chart in real time. Subsection 2-b shows using a keyboard shortcut to add an operation to an on-machine circuit layout.},width=\linewidth]{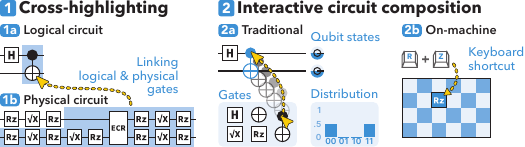}
    \caption{Interaction choice examples. (1) Cross-highlighting for exploration (from Patoka~\cite{kim2025:patoka}). (2a) Drag-and-drop composition with final qubit states and simulated outcome distributions (from IBM Composer~\cite{ibmComposer}. (2b) Keyboard shortcut-based circuit creation (from Crumble~\cite{crumble}).}
    \label{fig:interaction}
\end{figure}

\subsection{Interaction-level}

High-level interaction types include \textit{exploration} and \textit{composition}. 
For \feature{exploration}, prior approaches used \val{aggregation}, \val{filtering}, \val{brushing}, \val{tooltip} (\Cref{fig:running}-5), \val{cross-highlighting}, \val{slideshow}, and \val{toggling}.
For example, Quantivine~\cite{wen2024:quantivine} supports switching between aggregated gates and low-level gates and the provenance of a selected qubit by filtering out all the other qubits. 
Patoka~\cite{kim2025:patoka} supports cross-highlighting between a selected gate and its transpiled gates (typically more gates; \Cref{fig:interaction}-1), tooltip-based detail browsing, and self-paced animation (or slideshow) for an on-machine circuit (\Cref{fig:running}-2d). 
QuantumEyes~\cite{ruan2024:quantumeyes} allows for brush-selecting moments for browsing details (\eg~\Cref{fig:running}-5).
IBM's machine explorer (on-machine) shows gate and measurement error rates where users can toggle different error types.

For \feature{composition}, interactive tools commonly use \val{direct manipulation} approaches.
For example, IBM's circuit composer~\cite{ibmComposer} and Quirk composer~\cite{quirk} allow for dragging and dropping a gate from a panel to a traditional circuit format (\Cref{fig:interaction}-2a). 
For a small circuit, they have a preview of expected final quantum states and simulated measurement outcome distribution. 
QuFlow~\cite{lin2018:quflow} uses a \val{pop-up menu} to add gates.
Crumble~\cite{crumble} uses on-machine layout for circuit composition via \val{keyboard shortcuts} because its primary goal is simulating quantum errors and location information is critical for related problems (\Cref{fig:interaction}-2b). 
Crumble shows the time dimension using animation on the on-machine view along with a concatenated traditional circuit format.

%% file: sections/6-1-examples.tex
\section{Examples}\label{sec:examples}
We show how our design space captures quantum circuit visualization designs by describing three example cases in \Cref{fig:teaser}-A, D, and F.

\subsection{Traditional Circuit with Annotation}

\bpstart{View}
\Cref{fig:teaser}-A, reconstructed from Case \ex{7} and \ex{91}, shows a traditional circuit diagram.
This example uses a horizontally-directed traditional layout that maps qubits to one-dimensional vertical positions and moments (time) to horizontal positions. 

\bpstart{Component}
In addition to positions, qubits are represented as solid lines (bit lines) along with their indices (Q1, $\ldots$, Q3).
Gates are shown as widely known, conventional symbols, and the locations of those symbols mean applied qubits.
For example, the plus sign in a circle ($\bigoplus$) means a NOT gate (the first gate).
The other in-circle plus signs connected with a dot ($\bullet-\bigoplus$) are C-NOT gates (the second and third gates), where the dots are on control qubit lines and in-circle plus signs are on target qubit lines.
The gauge symbols ($\frown$\hspace{-3mm}$\nearrow$) in boxes indicate final measurements, and the vertical solid lines connect them to the classical register (the double line) to indicate where the measured outcomes are stored.
We note that there can be multiple classical registers (\#0, \#37) for complex processing purposes.

\bpstart{Management, and narrative}
The first three gates (in this case, three moments) are selected.
This selection is indicated using an emphasis (box) and annotation to describe their functional characteristics (multi-qubit entanglement).
The slash-symbol notation ($\neq$) on the double line for the classical register collapses two classical bits, with the number of classical bits annotated.

\subsection{Heatmap for Gate Fidelity}

\bpstart{View}
The example in \Cref{fig:teaser}-D (reconstructed from Case \ex{90}) concatenates two views.
Instead of circuit formats (direct representation of qubits and gates), this visualization uses a heatmap format to represent some quantitative values (\eg~error rates, decoherence time) regarding a physical quantum computer (on-machine). 
In this chart, the position encodings are already occupied by physical qubit locations.

\bpstart{Component, management, and narrative}
Unlike \Cref{fig:teaser}-C where each dot represents the corresponding qubits, this heatmap does not directly show qubits. 
Instead, each cell in the heatmap shows the connectivity between the corresponding qubit pair.
This is a common choice to show 2-qubit gate error rates (\eg~ECR gate, a physical gate for implementing C-NOT gate).
The color and stroke width of each cell encodes error rates (quantitative) and connectivity (or edge) type or groping (categorical). 
Each view has a title to show what it is about. 

\subsection{Animated Circuit}

\bpstart{View}
The animated on-machine, circuit diagram in \Cref{fig:teaser}-F is redrawn from Patoka's on-machine animation~\cite{kim2025:patoka}.
Each dot (or node) in a scene represents a qubit, and each edge indicates their connectivity.
Elements are positioned based on their locations on its physical chip. 

\bpstart{Component and management}
This example shows qubit connectivity, gate error on each qubit, and moment operation time.
First, the connectivity information matters for circuit optimization and error correction procedures.
Because not every pair of qubits, even if proximal to each other, is connected, showing it is important.
Next, each scene of this animation shows the error rate of the gate on each qubit (at the moment) by mapping it to the dot color.
Lastly, to show the time progression, this case maps the moment operation times (how long does it take to process a moment) to a time encoding.
Because physical operation times range at a nanosecond scale, it multiplies a constant factor to make the transition times long enough to be perceived.

\bpstart{Interaction} To allow for browsing each scene in detail, Patoka~\cite{kim2025:patoka}'s on-machine animation lets users pace it by themselves, resulting in an interactive slideshow.

%% file: sections/6-discussion-1.tex
\section{Discussion}

Our design space describes design choices in creating quantum circuit visualizations (RQ2). 
To better situate future research (RQ3), we reflect on our design space analysis with respect to popular visualization design principles.
Then, we identify future research directions in terms of data structure, scalability, and integrability. 

\subsection{Relationship to Prior Principles}

The following reflections aim to illuminate how visualization design principles can be applied to representing quantum circuits and point to missed opportunities that near-term future work could address.

\bpstart{Overview, detail, and complexity}
A circuit visualization can be highly complex with a few qubits and many gates.
It is difficult to present all the complex information within a quantum circuit while keeping every detail legible, motivating overview and detail presentation methods.
Relevant design principles include the \textit{Visual Information Seeking Mantra}~\cite{shneiderman2003:mantra} (``Overview first, zoom and filter, then details-on-demand'') and tradeoffs between screen space and abstraction in dashboard visualization~\cite{bach2023:dashboard}.

These principles assume detail-browsing interactions, such as tooltips, zoom, and coordinate filters. 
While there are a few interactive circuit visualization cases, interaction is certainly an underexplored area in quantum circuit visualization beyond tooltips~\cite{kim2025:patoka} and toggles~\cite{wen2024:quantivine}.
For example, quantum architecture research often looks at dynamics of a few thousands qubits, which requires the scalability of interaction methods.
More fundamentally, a formal modeling of interactions on quantum circuit visualizations is going to be necessary to enable diverse interaction methods in a flexible way.

Another related aspect is visual complexity.
The visualization community often strives to reduce visual complexity and clutter because they may harm and distort the user's perception~\cite{lo2022:misinformation}.
Among the clutter reduction approaches for statistical charts (Ellis and Dix~\cite{ellis2007:clutter}), consistency of spatial information, scalability, and ability to show attributes are relevant to quantum circuits with strict spatial constraints.
Specifically, focus and context techniques and pixel-plotting (converting data points into smaller visual units) are potentially applicable but less explored areas. 
Quantum circuit adds its unique nuances to exploring those complexity adjustment techniques.
Given a circuit with a few thousand gates, for instance, it will be challenging to indicate qubits affected by multi-qubit gates with the limited number of pixels.

\bpstart{Layout and perceptual effectiveness}
Quantum circuits have three core dimensions: 2-dimensional qubit locations on a machine and a time dimension. 
Given that 3-dimensional position encodings tend to be ineffective in perception~\cite{kjellin2008:2d3d,tittle1995:distortion,norman1996:length,todd2004:shape}, creators need to reason about this tradeoff of which information to prioritize for position encodings.
Thus, delivering such visual design principles to quantum community can be beneficial, for example, via declarative grammars and design recommenders, which is active in certain science domains like genomics~\cite{nusrat2019:genomic,lyi2022:gosling,pandey2023:genorec}.
For example, one can juxtapose traditional and on-machine layouts~\cite{crumble,kim2025:patoka,wen2024:quantivine}.
In doing so, future work should consider how to align novel techniques with conventions in the field given their wide usage in educational settings.

\bpstart{View consistency}
Qu and Hullman~\cite{qu2018:multiview} emphasize the consistency between multiple views in dashboard settings, such as axis and scales.
As quantum circuit visualizations can be composed of multiple views, multi-view consistency principles deem to be related.
For example, when there are multiple circuit formats (\eg~comparing logical and physical programs), keeping rows consistent to represent the same qubits would be an adaptation of the axis consistency constraint.
On the other hand, those constraints may not be suitable for cases with different layouts.
For example, future work can reason about intuitive cross-view mapping methods for the juxtaposition of on-machine and traditional layouts (as in Crumble~\cite{crumble}).

\bpstart{Collaborative visualization}
Working around quantum computers often involve collaboration, and collaborative visualization for quantum circuits lacks software support.
Isenberg~\ea~\cite{isenberg2011:collaborative} emphasize that collaborative visualization should aim for ``an increased understanding'' into a given decision space. 
In quantum computing, we are aware of two scenarios for collaboration: (1) a shared repository of programs (\eg~Quantum Algorithm Zoo\footnote{\url{https://quantumalgorithmzoo.org/}}, Quantum Advantage Tracker\footnote{\url{https://quantum-advantage-tracker.github.io/}}) and (2) a team of practitioners. 
First, repositories could adopt insights from asynchronous communication support, such as sharing ideas and producing further discussion~\cite{viegas2007:manyeyes} and using annotations as comments to enhance clarity~\cite{heer2009:vv}.
Systems for collaborative circuit visualizations should consider diverse selection units and methods as our findings indicate.
Second, it will be important to preserve the provenance and log data of a quantum circuit, given that collaborations are often done via source codes (in notebooks) shared using version control systems like Git.
In doing so, future systems must address a dimensionality challenge as displaying the evolution of a circuit adds another dimension to an already high-dimensional space.

%% file: sections/7-discussion-2.tex
\subsection{New Research Directions for Quantum Circuit Visualization Systems}

A higher-level purpose of our design space analysis is to offer guidance for next-generation quantum circuit visualization systems that are well integrated with quantum programming tasks across different platforms.
Our blueprint includes modeling data structures, managing scalability, and integrability with existing software ecosystems. 
Solving these challenges will require interdisciplinary efforts, not only among visualization, CHI, and quantum computing communities but also including programming language, database, computer science education, and beyond.
In our discussion, we assume a grammar-based approach, as it has powered numerous effective visualization systems across various domains and platforms.

\bpstart{Data models for quantum circuit visualizations}
Many visualization grammars have an assumption about supported data models
(\eg~tabular data for ggplot2~\cite{hadley2010:ggplot2} and Vega-Lite~\cite{satyanarayan2016:reactive}, node-link structures for NetworkX~\cite{networkx}). 
Current quantum circuit visualization tools rely on how a given hardware provider models quantum programs, and they are not necessarily consistent.
Although QASM~\cite{qasm} is considered a standard approach to representing quantum programs, it cannot express other key information, such as qubit layout, fidelity, narrative elements, and so on.
Thus, visualization data models need dedicated efforts to build platform-agnostic quantum circuit visualization toolkits on top of conventions widely used in the field.

To motivate future research on data models for circuit visualizations, we suggest the following guidelines in addition to cross-platform feasibility.
First, they should be able to accommodate both quantum programs and visualization needs, such as capturing circuits at varying degrees of abstraction for narrative and interactivity.
Next, such data models should not be difficult to use. 
For example, compiling quantum program source codes into such a data model will improve the usability of circuit visualization systems greatly.
Because data for circuit visualizations are essentially programs, modeling circuit data would require broader interdisciplinary collaboration beyond visualization and quantum computing. 
For example, programming language-based approaches could be useful for defining type systems for circuit components and their structures.
Database-inspired interactive query methods could support incorporating data models and interactions scalably as quantum circuit visualizations can vary in size (\Cref{fig:dist}-C).

\bpstart{Cognition}
Quantum circuit visualizations involve some kind of aggregation or smoothing methods.
As semantically categorizing data is important for aggregation in statistical charts~\cite{setlur2022:oscar}, ideally, circuit visualization techniques should also aim to improve user reasoning about quantum programs.
Computer science education research on different program paradigms have emphasized how people mentally conceptualize programs.
Mental models are not just the functional structures of different components, but also include the relationship of components to domain-related problems~\cite{corritore1999:mental}, which our design space indicates with respect to the use of attachments.
Quantum computing has diverse, specialized application areas, including chemistry, stock portfolio, or logistics optimization problems.
Thus, future research can consider mental models of quantum programs in relation to diverse domain areas, to help researchers formulate semantic units of quantum program components and aid learners in reasoning about them.

\bpstart{Task-effective integration with ecosystems}
Quantum circuit visualizations are often used within computational notebooks~\cite{ashktorab2019:hqci}. Yet, our analysis shows that existing toolkits provide limited support for narrative features, which are often added later using graphic tools.
Thus, future tools for scalable, interactive, and narrative-driven quantum circuit visualization will need to be developed and integrated into existing software ecosystems.
To support this integration, we need a deeper understanding of the tasks and needs of practitioners.
For example, beginners would find interactive circuit composers to be useful, whereas manually dragging thousands of gates would hamper experts.
Extending high-level purposes discussed in our design space, a more fine-grained task taxonomy would be beneficial for designing toolkits as well as evaluating them.
Future research must consider common and unique aspects of quantum circuit visualization tasks in relation to various visualization task models (\eg~Brehmer~\ea~\cite{brehmer2013:multilevel}, Amar~\ea~\cite{amar2005:task}).

\section{Conclusion}
To pave paths for future quantum circuit visualization systems, we curated a corpus of 182 static and 12 interactive cases with detailed open-codes and analyzed them into a design space. 
Our design space characterizes quantum circuit visualizations as design choices at view, component, management, narrative, and interaction levels. 
We reflected on our design space with popular visualization design principles, including overview+detail, encoding effectiveness, multiple view consistency, and collaborative visualizations, to capture missed opportunities.
We further discuss technical research opportunities for quantum circuit visualizations in terms of data models, cognition, and integrability.
We proivde a design gallery of quantum circuit visualizations via \url{https://see-mike-out.github.io/qc-circ-gallery/}.

\bpstart{Limitation} 
Our work only looks at circuit visualizations for gate-based quantum computing given that it is a primary method with commercial availability and widely shared technical stacks.
Future work can expand our work by looking at other major quantum computing methods, such as photonic analogue quantum computing. 
Next, the design strategies shown in our work do not represent ``good'' designs but provide a snapshot based on the corpus curated by us. 
To better quantum circuit visualization tools, future work should aim to understand task effectiveness of different representation and interaction techniques.